\documentclass[fleqn,twoside]{article}
\usepackage{espcrc2}

\usepackage{graphicx}

\newcommand{\AmS}{{\protect\the\textfont2
  A\kern-.1667em\lower.5ex\hbox{M}\kern-.125emS}}

\title{Study of the Finite Density State based on SU(2) Lattice QCD}

\author{Shin Muroya\address{Tokuyama Women's Coll. Tokuyama, 745-8511, Japan},
        Atsushi Nakamura\address[IMC]{IMC, Hiroshima University,  
        Higashi-Hiroshima 739-8521,  Japan }
        and
        Chiho Nonaka\address{Radiation Lab, RIKEN, 2-1 Hirosawa, Wako, Saitama, 351-0198, Japan}\thanks{Presenter}
        }
       
\begin{document}

\begin{abstract}
We report our recent numerical studies on two-color QCD with Wilson fermions.
First the phase structure of the system in $(\kappa,\mu)$-plane
is analyzed by measuring Polyakov line, baryon number density, gluon
energy density and Polyakov line susceptibility.
Then meson and baryon (di-quark) propagators are investigated.
We find that the vector meson mass decreases near $\mu_c$.
This phenomena can be related to the low-mass lepton-pair 
enhancement observed in CERES experiment.
Since our lattice is small, in order to confirm the results, 
we calculate both periodic
and anti-periodic boundary condition cases. 
Preliminary calculations of 
color averaged, symmetric and anti-symmetric forces of $qq$ system 
obtained from Polyakov line correlations, and gluon propagators
are reported.
\vspace{-1pc}
\end{abstract}

\maketitle

\section{Introduction}

Finite density QCD has attracted much attention recently:
we are expecting that QCD has a rich structure when we add a density
or chemical potential axis in parameter space. 
The finite temperature deconfinement transition line starts at $\mu=0$,
which might be a cross-over accompanied by a tri-critical point at finite
$\mu$.
At high $\mu$ and relatively low temperature, there would be 
a new form of matter 
such as color superconductive phase, or color-flavor locking phase, which are
speculated by mainly perturbative calculations \cite{CCC}.  

Non-perturbative study of the finite density QCD by lattice simulations
is highly required, but because of the famous difficulty, the direct
Monte Carlo simulation of SU(3) system is very difficult.
Many approaches have been reported, 
\cite{kaczmarek,Taro01,Fodor,Ejiri,Philipsen},
and yet, no reliable algorithm is known to investigate color SU(3) QCD at
low temperature with large chemical potential.
Color-SU(2) is currently the unique model by which we can study these 
regions non-perturbatively.
After the first trial of the calculation\cite{Nakamura84}, there had
been no work during many years.  
Because of the recent rapid progress of theoretical 
understanding of the system \cite{Kogut00,KST}, however, 
the calculation has been activated \cite{Lombardo99a,Hands99,Morrison}.

In this report, we study the finite density QCD by Wilson fermions.
We investigate the phase structure in $(\kappa,\mu)$-plane for
Iwasaki improved-gauge action. 
In order to take account of the fermion determinant,
we use an algorithm in which the ratio of the determinant
is evaluated explicitly at each Metropolis update process \cite{MNN00}.

\section{Phase Structure}

\begin{figure}
\begin{center}
\includegraphics[width=.9 \linewidth]{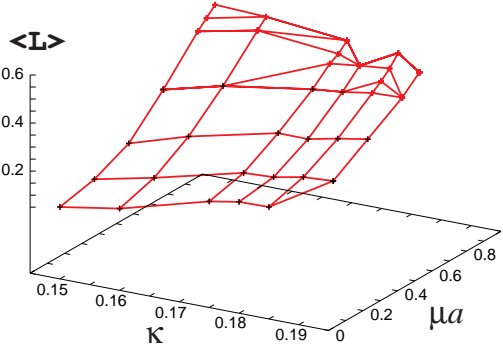}
\end{center}
\vspace{-13mm}
\caption{Behavior of Polyakov line in $(\kappa,\mu)$-space.}
\label{Fig-Pol}
\end{figure}

In Fig.\ref{Fig-Pol}, we show 
the average 
of Polyakov line
in $(\kappa,\mu)$-plane at $\beta=0.7$.  We evaluate also
baryon number density, gluon energy density and Polyakov line susceptibility,
and especially from the peak of the Polyakov line susceptibility 
we can determine the confinement/deconfinement phase transition regions.
In both bounday conditions\footnote
{Periodic (anti-periodic) refers the periodic
(anti-periodic) boundary condition for quark fields in spatial
directions.}, 
the behaviors of these quantities are essentially
the same \cite{MNN00}.

We study several cases by varying lattice size 
($4^4$, $4^3\times8$, $4^3\times12$),
the boundary conditions and the number of flavor ($N_f=2, 3$).

Calculations become unstable and break down 
at large chemical potential regions\cite{MNN00}.
In such cases, negative determinant appears
for $N_f=3$.

\section{Meson and diquark propagators}

\begin{figure}
\begin{center}
\includegraphics[width=.9 \linewidth]{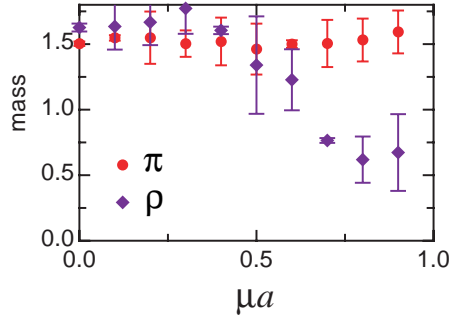}
\end{center}
\vspace{-15mm}
\caption{$\pi$ and $\rho$ masses as a function of $\mu$ for $\kappa=0.160$
with periodic boundary condition}
\label{Fig-mass160P}
\end{figure}

We calculate meson and di-quark propagators in pseudo-scalar and vector
channels.
We find that $\rho$ meson mass drops as $\mu$ reaches to the
critical region.  In order to confirm this unexpected result, we
calculate both periodic and anti-periodic boundary conditions and
several $\kappa$'s ($\kappa=0.150, 0.160, 0.175$).
In Fig.\ref{Fig-mass160P} 
we show $\pi$ and $\rho$ masses as a function of $\mu$
for $\kappa=0.160$ (periodic boundary condition). 
Although our statistic is not enough to extrapolate them to the chiral
limit, the signal of the anomalous behavior of the vector channel 
is clearly seen.

At $\mu=0$, because of the QCD inequality \cite{KST},
the pseudo-scalar channel has the lightest mass, but
there is no such a condition for the finite density state.
Indeed there are several conjectures in literatures \cite{rho-theo}.  
%
If our result shown in Fig.\ref{Fig-mass160P} is not a special feature
of 
the color SU(2) model, this is the first lattice
observation of the drop of the vector meson mass which may play an important
role in the phenomenological analyses on large low-mass lepton-pair 
enhancement measured in CERES \cite{CERES}.

Another possibility is that the vector meson channel is mixed with
the vector diquark one, $\psi^{c}\gamma_5\gamma_\mu\psi$, at high $\mu$
and their masses are distorted\footnote{We thank Thomas Schaefer for
useful discussions on 
$\bar{\psi}\gamma_\mu\psi$/$\psi^{c}\gamma_5\gamma_\mu\psi$ mixing .}.
In this case, we can study ``meson" and ``diquark" mixing in color
SU(2) QCD.

\section{Force between $qq$ / $\bar{q}q$}

In the previous section, we show that the vector meson may have
a peculiar behavior at finite density.  Above $\mu_c$ at low
temperature, 
the condensation of $qq$ may occur. 
We need tools to understand the vacuum structure numerically.


From Polyakov line correlations, color average heavy quark potential
is calculated. 
One can also obtain color singlet ($V_1$) and triplet potential ($V_3$) for
$q\bar{q}$ system: 
$N_c\times\bar{N_c} = 1 \oplus (N_c^2-1)$ \cite{Nadkarni}.
Color symmetric ($V_s$) and anti-symmetric ($V_a$) potential for
$qq$ system can be also evaluated: 
$N_c\times N_c = \frac{1}{2}N_c(N_c+1) \oplus \frac{1}{2}N_c(N_c-1)$. 
In color SU(2), $V_1=V_a$ and $V_3=V_s$ even at $\mu \ne 0$.
Therefore if there would exist difference between $qq$ and $q\bar{q}$ states,
it should originate in Dirac structure.

\begin{figure}
\begin{center}
\includegraphics[width=.9 \linewidth]{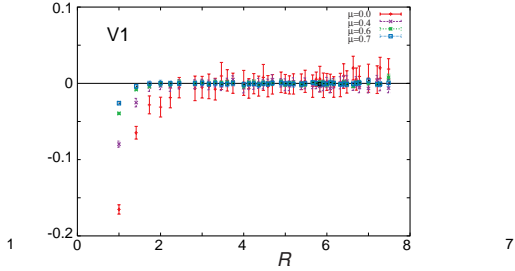}
\end{center}
\vspace{-10mm}
\caption{Color singlet potential $V_1$ for $\mu=0.0, 0.4, 0.6$ and 0.7.}
\label{Fig-V1}
\end{figure}

In Fig.\ref{Fig-V1}, we plot preliminary data of color singlet 
potential. 
The rotational invariance is good by virtue of the renormalization
group improved gauge action.
The singlet (triplet) potential is attractive (repulsive) and
the forces become weaker as $\mu$ increases.
We fix the gauge in Lorentz gauge for the color-dependent
force and gluon propagators discussed in the following.



We study gluon propagators as a tool to
investigate the finite density QCD,
\begin{eqnarray}
G_{\mu\nu}(P_0,P_x,P_y,z)
= \langle \mbox{Tr}
A_\mu(P_0,P_x,P_y,z)          \nonumber \\
\times
A_\nu(-P_0,-P_x,-P_y,0) \rangle .
\end{eqnarray} \noindent
We set the momenta as $(P_0,P_x,P_y) = (0,2\pi/N_x,0)$.
The electric and magnetic parts of the propagator, 
$G_e = G_{tt}$ and $G_m = G_{yy}$, are expected to 
behave as,
$
G_{e(m)}(z) \propto \exp(-E_{e(m)}z).
$
When the system is screened at the finite temperature and/or
finite density, 
the screening effects should appear as 
$E_e = \sqrt{(P_x^2+M_e^2)}$ and $E_m = \sqrt{(P_x^2+M_m^2)}$, 
with $M_e$ and $M_m$ being electric and magnetic masses of
the gluons, respectively.
At $T=0$ and $\mu=0$ the screening is infinite and the propagators
vanish at long distance \cite{nakamura98}.

\begin{figure}
\includegraphics[width=.85  \linewidth]{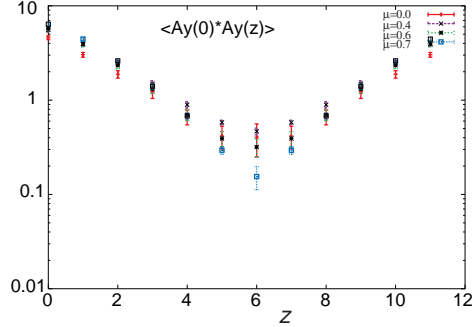}
\vspace{-10mm}
\caption{Transverse gluon propagator for different $\mu$.
Lattice size is $8\times 4\times 12\times 4$.
}
\label{Fig-Gyy}
\end{figure}

We show in Fig.\ref{Fig-Gyy} a preliminary result for $G_m(z)$.
As $\mu$ increases, the screening becomes weaker, i.e., the system
shows the deconfinement feature, but at $\mu=0.7$ the propagator
at long range seems to drop.  It is an urgent future task to
confirm the phenomena, because it could be an indication of
a new phase at large $\mu$. 

\noindent
{\bf Acknowledgment}
This work is partially supported by Grant-in-Aide for Scientific Research by
Monbu-Kagaku-sho (No.11440080 and No. 12554008) and ERI of Tokuyama Univ.
Simulations were performed on SR8000 at IMC, Hiroshima
Univ., SX5 at RCNP, Osaka Univ., SR8000 at KEK and
VPP5000 at Science Information Processing Center, Tsukuba Univ.


\end{document}